# A narrow-linewidth external cavity quantum dot laser for high-resolution spectroscopy in the near-infrared and yellow spectral ranges

A. Yu. Nevsky*, U. Bressel, Ch. Eisele, M. Okhapkin, S. Schiller
*Institut für Experimentalphysik, Heinrich-Heine Universität Düsseldorf, Germany*
www.exphy.uni-duesseldorf.de

A. Gubenko, D. Livshits, S. Mikhrin, I. Krestnikov, A. Kovsh
*Innolume GmbH, Dortmund, Germany*
www.innolume.com

* *E-mail*: alexander.nevsky@uni-duesseldorf.de

We demonstrate a diode laser system which is suitable for high-resolution spectroscopy in the 1.2 µm and yellow spectral ranges. It is based on a two-facet quantum dot chip in a Littrow-type external cavity configuration. The laser is tunable in the range 1125 - 1280 nm, with an output power of more than 200 mW and exhibits a free-running linewidth of 200 kHz. Amplitude and frequency noise were characterized, including the dependence of frequency noise on the cavity length. Frequency stabilization to a high-finesse reference cavity is demonstrated reducing the linewidth to about 20 – 30 kHz. Yellow light (> 3 mW) at 578 nm was generated by frequency doubling in an enhancement cavity containing a PPLN crystal. The source has potential application for precision spectroscopy of ultra-cold Yb atoms and molecular hydrogen ions.

## 1. Introduction

Single-mode external cavity grating stabilized diode lasers (ECDL) are well established for a wide variety of applications thanks to their compact size, relatively low cost, large wavelength tuning range (of the order of 100 nm for laser diodes with antireflection coating), high output power and reliable operation. Depending on the materials and technology used, quantum well (QW) lasers can be manufactured for different wavelength ranges extending from the blue to the mid infrared part of the spectrum (Fig.1)

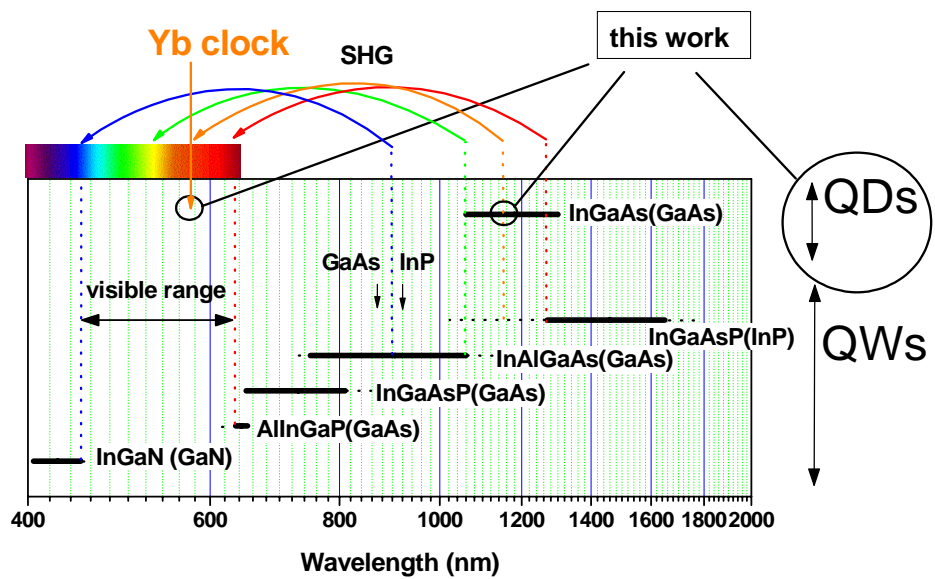

**Fig. 1.** Wavelength coverage of laser diodes of different types and materials.
QD: quantum dot lasers, QW: quantum well lasers.

However, some spectral ranges are impossible or difficult to reach because of material limitations. For example, the occurrence of strain-induced dislocations in InGaAs QW lasers limits the longest lasing wavelength to approx. 1.1 µm. Low electron barriers, strong Auger recombination, and insufficient refractive index contrast do not allow creation of high-performance InGaAsP lasers with wavelength shorter than aprrox. 1.3 µm. InGaAs quantum dot (QD) lasers, invented in the early nineties [1], are filling the gap between 1100 nm and 1300 nm, opening new perspectives for a variety of applications. In particular, the wavelength range covered by their second-harmonic radiation lies in the yellow. While in this range tunable, narrow-linewidth radiation is available from dye lasers, continuous-wave optical parametric oscillators [2], or sum-frequency generation [3], clearly a frequency-doubled diode laser would represent a strong improvement in terms of cost, complexity and occupied volume. Compared to the well-developed external cavity QW lasers, where already sub-Hz laser linewidths by active stabilization to reference resonators have been demonstrated [4-6], the current state-of-the-art of QD lasers is at a very early stage. Recently, the coherence properties of QD lasers under external feedback conditions were investigated [7], and linewidths of the order of several GHz were demonstrated. Here we present a detailed study of a QD laser in the external cavity Littrow configuration and demonstrate for the first time, that these lasers are in principle suitable for high-resolution experiments, with linewidth well below the 100 kHz level. We focus on the particular wavelength of 1156 nm, since the second harmonic of this radiation at 578 nm matches the clock transition in the neutral ytterbium atom, a candidate for an optical frequency standard with less than $10^{-16}$ inaccuracy [8].

## 2. Quantum dot laser chip

The semiconductor structure for the gain chip was grown by molecular-beam epitaxy. A 0.5µm-thick GaAs waveguide is confined by 1.5µm-thick $Al_{0.35}Ga_{0.65}As$ claddings doped with Si and C for n- and p-type conductivity, respectively. Quantum dots were formed by deposition of 0.8 nm-thick InAs insertion covered with an $In_{0.15}Ga_{0.85}As$ cap of variable thickness and a 33 nm thick GaAs spacer. A gain region consists of 10 non-identical QD planes designed for continuous tuning between GS and ES optical transitions [9]. A similar QD gain region has recently been used to demonstrate a lasing spectrum as broad as 75 nm [10]. A 5-µm-wide bent ridge with a normal output facet and 5°- tilted rear facet was fabricated using a standard photolithography process, reactive ion etching and cleavage. The facets were anti-reflection coated with a residual reflectivity of ~ 0.5%. The gain chips are mounted p-side up on the AlN carriers.

## 3. General properties of the QD-ECDL

The schematic of the QD-ECDL is shown in Fig. 2. The radiation emitted from the tilted facet (angle of 17° with respect to the normal to the waveguide) is collimated with an AR coated aspheric lens.

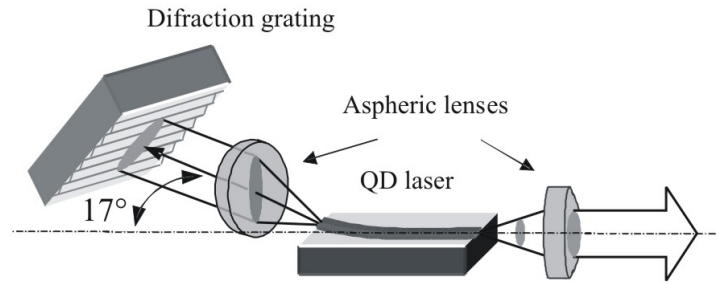

**Fig.2.** Schematic of the external cavity quantum dot laser.

A diffraction grating reflects the first diffraction order back to the laser chip, forming a Littrow-like configuration. Coarse wavelength tuning of the laser is obtained by changing the incidence angle of the grating. For fine tuning of the laser frequency a piezo transducer (PZT) that displaces the grating is used, allowing a mode-hope-free tuning range of about 3 GHz. The output radiation of the laser is also collimated by an aspheric lens and corrected by an AR coated prism pair, forming an almost square cross-section beam of size approx. 6 x 6 mm. The whole construction including the laser chip, objectives and the grating is mounted on a copper plate, the temperature of which is stabilized with Peltier elements to better than 1 mK. The laser is driven by a custom made ultra-low noise current source (max output current 750 mA) with a residual RMS current noise on the order of 1 µA. The typical output power of the laser at the wavelength 1156 nm as a function of the operating current is shown in Fig. 3.

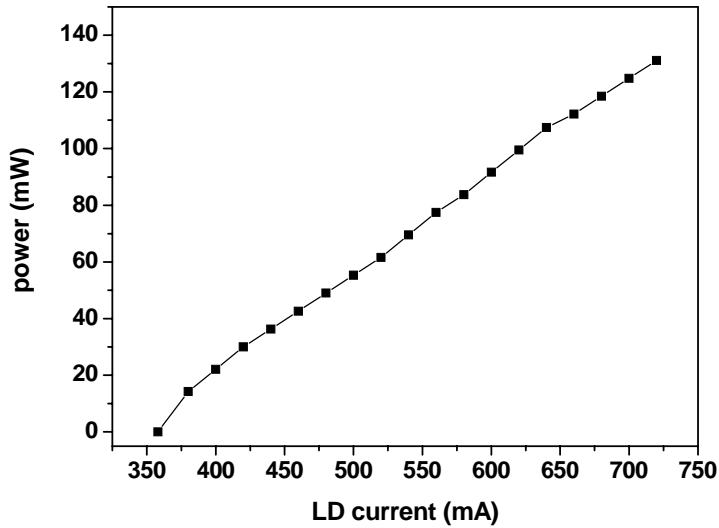

**Fig. 3.** Output power of the QD-ECDL at 1156 nm as a function of the operating current.

In this setup ("low-noise" implementation) the laser wavelength could be changed in the range from approx. 1148 nm to 1250 nm just by tilting the grating. Under proper alignment of the reflection grating the sidemode suppression was measured to be more than 35 dB within the whole emission range. Using operating currents above 750 mA, a tuning range exceeding 200 nm as well as an output power of more than 500 mW at a central wavelength of about 1180 nm has been achieved (see Fig.4) still keeping single spatial mode. This range is comparable to but the power is at least one order higher than for the externally tunable InGaAs QD device reported earlier [11].

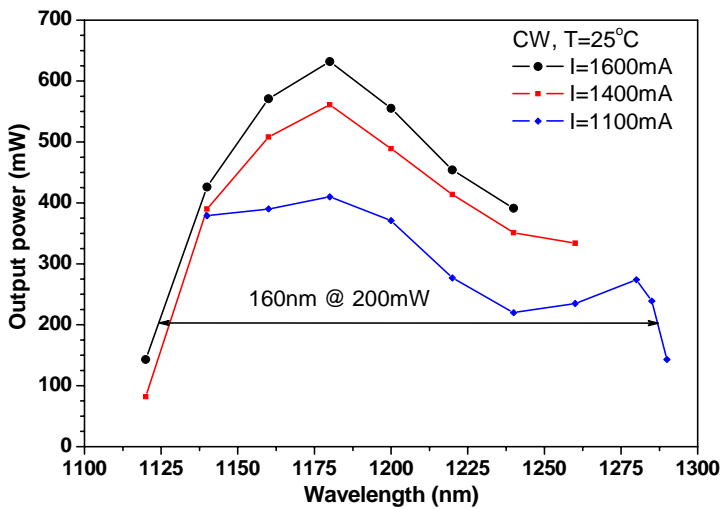

**Fig. 4.** Tunability and output power of the QD-ECDL under high operating currents.

## 4. Linewidth and frequency noise

The free-running linewidth of the QD-ECDL (Fig. 5) was measured by producing the heterodyne beat between two almost identical laser systems, tuned to to 1165 nm. The beat was detected with a high-bandwidth photodetector and analyzed with a spectrum analyzer.

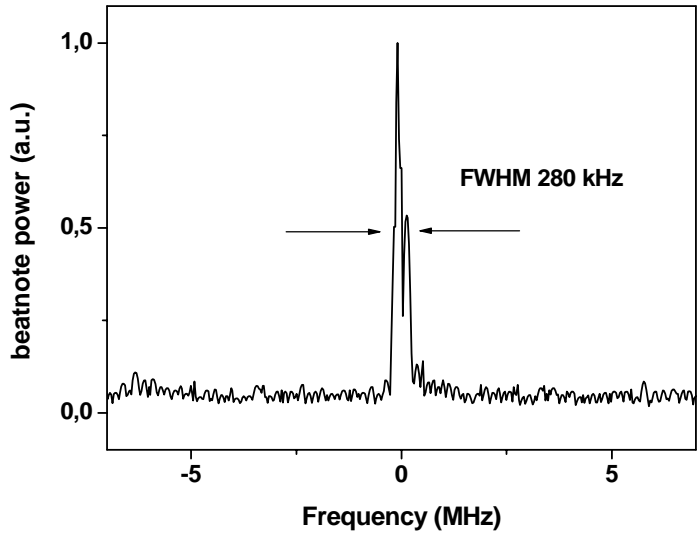

**Fig. 5.** Power spectrum of the beat note between two free-running QD-ECDLs (cavity lengths approx. 35 mm). The spectrum analyser resolution bandwidth is 30 kHz.

The observed beat linewidth of 280 kHz (on a timescale of several 10 µs) implies a linewidth $\sqrt{2}$ smaller, i.e. 200 kHz for a single laser. More detailed analysis of the laser linewidth was not possible with this method due to a large frequency jitter of several MHz on a 1 s timescale, caused probably by current source noise and mechanical vibrations. In order to perform a detailed analysis of the laser frequency noise, in particular of the intrinsic linewidth, we used an external ring cavity as a frequency discriminator (Fig. 6-a).

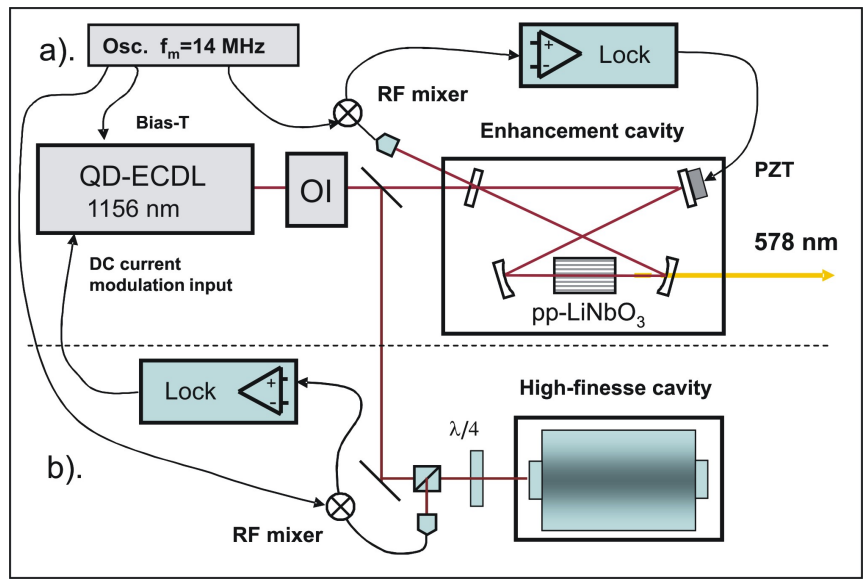

**Fig. 6.** Frequency stabilization and frequency doubling setup (OI – optical isolator, pp: periodically poled). a). Setup for frequency noise characterization and for SHG in an enhancement cavity containing a pp-LiNbO$_3$ crystal. The cavity is locked to the laser. b). Additional setup for frequency stabilization of the laser to the high-finesse cavity.

This cavity is the same used later for resonant frequency doubling to the yellow spectral range (see Sec.6). The resonator is formed by four mirrors, has a 750 MHz free spectral range, and possesses a finesse of about 200 at the wavelength 1156 nm (including the nonlinear crystal). One of the cavity mirrors is mounted on a piezo actuator (PZT), allowing to tune the cavity resonance frequency. The cavity is stabilized to the laser frequency using the Pound-Drever-Hall (PDH) technique. Modulation sidebands at 14 MHz were generated by modulation of the injection current of the laser, using a bias-T implemented in the current controller. In order to prevent feedback from the cavity on the QD-ECDL frequency, two optical isolators (OI) in series were required. We used two available standard isolators optimized for 1064 nm, thus introducing high transmission losses at the wavelength of 1156 nm. The laser wave reflected from the cavity was detected by a fast photo-detector and demodulated using a double-balanced RF mixer. The cavity was locked to the laser using the PZT of the cavity mirror. The servo bandwidth was several kHz, nevertheless providing a stable lock over several hours.

The frequency fluctuations of the laser were analyzed from the closed loop error signal at the output of the frequency mixer. The bandwidth of this signal was about 7 MHz. For frequencies above the locking bandwidth (approx. 10 kHz), the fluctuations correspond to those of the unlocked (unstabilized) laser. The power spectral density of the frequency fluctuations is shown in Fig.7.

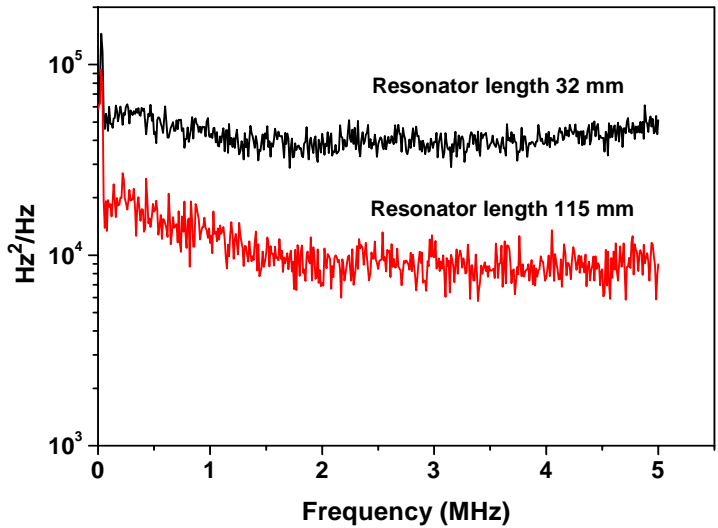

**Fig.7.** Power spectral density of the QD-ECDL frequency fluctuations for two different lengths of the laser resonator. The spectrum is corrected for the effects of the enhancement cavity bandwidth (3 MHz FWHM).

The frequency noise was characterized for two distinct QD-ECDL configurations, having resonator lengths of 32 mm and 115 mm, respectively, from the grating to the output AR facet of the laser chip. The larger resonator length showed a reduced sensitivity of the laser frequency to the operating current (30 MHz/mA in comparison to 100 MHz/mA for the 32 mm resonator). This also led to a factor 9 reduction of the frequency noise level (see Fig.8). A simple explanation for this is the reduced influence of the optical path length fluctuations occurring in the laser chip when the total cavity length increases.

Fluctuations of the optical path in the laser diode chip can also be caused by the instability (noise) of the operating current. To estimate this effect, the laser was connected to a 12 V car battery simply by adding a resistor in series to the laser, thus forming a simple current source without any active elements. Johnson noise in the series resistor (about 20 Ohm) leads to an RMS current noise of about 0.1 μA, about 10 times less than that of the laser diode driver. However, no difference in the laser frequency noise in comparison to the laser diode driver was found for either resonator length. Thus, the observed noise level is intrinsic to the lasing process.

Assuming white frequency noise, the "fundamental" laser linewidth $\Delta \nu_L$, can be estimated using the expression [12]:

$$\Delta \nu_L = \frac{\pi (\delta \nu_{rms})^2}{B} = \pi S_\nu \qquad (1)$$

where $\delta \nu_{rms}$ is the RMS frequency noise of the laser, $B$ is the bandwidth and $S_\nu$ is the power spectral density of the frequency noise in Hz$^2$/Hz. Using the frequency noise level from the high-frequency part of Fig.7, Eq.(1) yields a laser linewidth of about 180 kHz and 60 kHz for the short and long resonators, respectively. This simple analysis does not take into account the low-frequency instability of the laser frequency, caused by mechanical instability of the laser resonator, mains pick-up and other similar parasitic effects which result in a larger linewidth.

## 5. Frequency-stabilization to a high-finesse reference cavity

In order to reduce the laser linewidth and frequency drift, frequency stabilization to a reference cavity was implemented (see Fig.2b). After the optical isolators, a part of the beams is split off with a beamsplitter and coupled to the high-finesse reference cavity. As a reference cavity we used a sapphire Fabry-Perot resonator, of the type used for cryogenic relativity experiments at 1064 nm [13]. The resonator consists of a 30 mm long sapphire cylinder with a diameter of 25 mm and two high-finesse mirrors, optically contacted to its end faces. At the wavelength of 1156 nm the finesse of the cavity is about 10 000. The cavity is placed inside a vacuum chamber and temperature stabilized with a Peltier element to better than 1 mK. The temperature sensitivity of the cavity was measured to be about 100 MHz/K (thermal expansion coefficient 3.8 x $10^{-7}$/K).

The laser radiation reflected from the cavity, is detected with a low-noise photodetector. The PDH signal is obtained in the standard way. The PID type servo system controls the laser diode current. A stable lock of the laser to the high-finesse cavity was obtained. The bandwidth of the lock is about 500 kHz with a 1/f roll-off at low frequencies. To characterize the locking quality and residual frequency fluctuations, the enhancement cavity was again used as a discriminator. The cavity was also locked to the laser and the error signal of this lock was measured with a low-frequency FFT spectrum analyzer. The frequency noise spectrum both in free-running and cavity-stabilized operation are shown in Fig.8. Note that the frequencies below several kHz can not be taken for an analysis since they are suppressed by the locking system of the enhancement cavity. For comparison, the frequency fluctuations of a free running diode-pumped Nd:YAG laser at 1064 nm with intracavity doubling to 532 nm [14] are also shown. For this laser, the measurements were performed in the same way as with the QD-ECDL, i.e. the enhancement cavity was locked to the 1064 nm laser. At frequencies above several 10 kHz, the free-running frequency noise of the Nd:YAG laser is a factor 100 smaller than that of the long-cavity QD-ECDL. From Fig.8, and using again Eq.(1), the residual QD-ECDL linewidth can be estimated to be on the level of several 10 kHz. Due to the relatively small modulation index, the higher frequencies, not influenced by the locking system, contribute only to the phase noise of the laser. They should be removable by transmitting the radiation through another narrow-linewidth cavity.

A more broadband lock of the QD-ECDL as well as a better discriminator slope (a high-finesse cavity with a narrower transmission peak) are necessary in order to further reduce the laser linewidth.

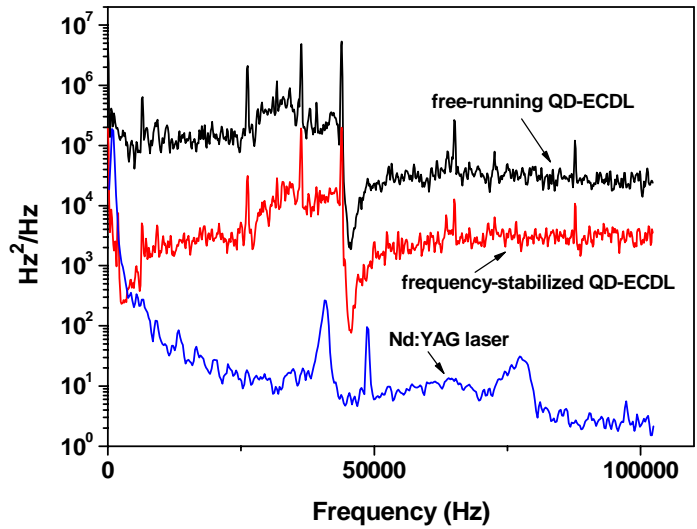

**Fig.8.** Power spectral density of frequency fluctuations of the free running and frequency stabilized QD-ECDL, obtained from the error signal of the enhancement cavity locked to the laser. The peaks at about 45 kHz are probably spurious signals. Note that the resonator lock bandwidth is about 10 kHz, thus the frequency fluctuations in this range are not representative for the free-running laser properties. The lower curve is a PSD of the free-running Nd:YAG laser at 1064 nm.

## 6. Second-harmonic generation

Second harmonic generation (SHG) is performed in an external enhancement cavity containing a periodically poled $LiNbO_3$ crystal (PPLN) (Fig.6-a). The fundamental wave of the QD-ECDL is resonantly enhanced in the ring cavity employed and described above. The finesse of the cavity is about 200. The PPLN crystal is 25 mm long, 0.5 mm thick and possesses a grating period of 8.33 μm. The crystal end faces are AR coated at the fundamental and harmonic wavelengths. To prevent optical damage or photorefractive effects, the PPLN crystal is operated in a small oven. At a temperature of about 190°C phase matching for doubling of 1156 nm radiation occurs. The thermal tuning coefficient of the phase-matched fundamental wavelength is 1.15 nm /K. The laser radiation is mode-matched to the cavity using a single focusing lens. A coupling efficiency of ~ 50% is obtained, due to non-perfect laser beam shape and non-optimized input coupler mirror. The two isolators transmitted only a fraction of the diode light, resulting in about 32 mW available before the enhancement cavity. Further attempts to increase the incident power by adjusting the optical isolators led to an increased feedback and, as a result, to an unstable lock of the cavity. The power of the generated second-harmonic wave as a function of the incident power is presented in Fig. 9. The dependence is nearly quadratic, indicating that fundamental wave depletion effects are not yet present at these power levels.

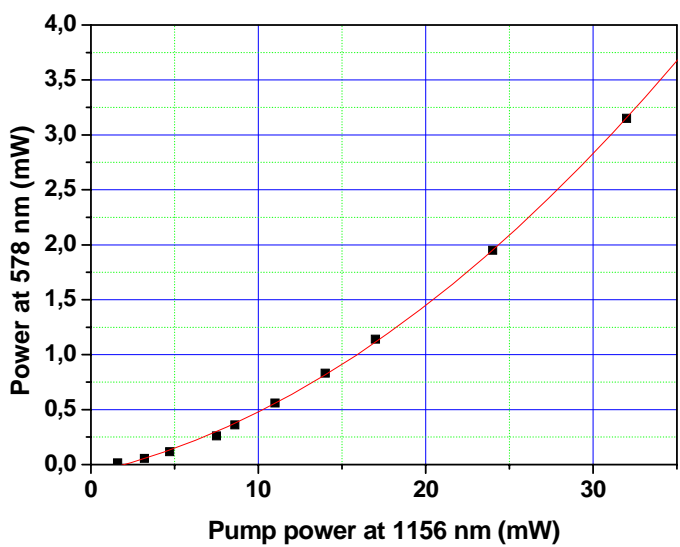

**Fig. 9.** Second harmonic power (578 nm) emitted from the doubling cavity as a function of the fundamental power at the input of the cavity. Smooth line is quadratic fit to the data.

Note that these results were achieved with a non-optimal input coupling mirror. We estimate that with an optimized input coupler transmission, the SHG power could be increased by a factor 2. In addition, the use of better optics (optical isolators, etc.) will allow to increase the in-coupled laser power, so that a few tens mW power in the yellow should be achievable.

## 7. Summary and outlook

We have shown that the quantum-dot external cavity diode laser has properties comparable to those of conventional external cavity diode lasers based on quantum wells. The free-running spectral properties make the laser suitable for standard absorption or fluorescence spectroscopy, including Doppler-free spectroscopy, with spectral resolution at the several MHz level. Since, as we demonstrated, frequency locking to a reference cavity and linewidth reduction to several 10 kHz is possible, QD-ECDLs are also suitable for spectroscopy at much higher resolution. We demonstrated these features at 1156 nm, but the same properties are expected at the second harmonic and at other wavelengths within the tuning range 1150-1280 nm, thus making the QD-ECDL a versatile source for high-resolution experiments.

Two applications in the field of ultra-high resolution spectroscopy are expected to be facilitated by this source. The first is an optical atomic clock in ytterbium [8] whose ultra-narrow clock transition lies at 578 nm. For this application, a power of ~1 mW at 578 nm is sufficient. While the clock radiation can be implemented also using a source based on sum frequency generation, a diode laser-based clock laser is of significant interest for cost reasons or once such a clock is to be made compact or even space-suitable. A second application are vibrational transitions in the molecular hydrogen ion $HD^+$, a molecule of interest for tests of quantum electrodynamics and for an alternative determination of the electron-to-proton mass ratio [14]. The $v = 0 \rightarrow v = 5$ vibrational overtone transition has a rotational substructure with 10 transitions in the range 1149 nm – 1201 nm [16], whose natural linewidths are 11 Hz [17]. A power level of a few mW is sufficient for the spectroscopy. In both applications, a small fraction of the near-infrared light (~ mW level) is required for optical frequency measurement by a frequency comb.

Thus, laser linewidths below the 1 and 10 Hz level, respectively, are desirable for these applications. Such a level should be achievable with the present source. For example, the achieved stabilization to the finesse 10 000 - cavity could be used as prestabilization step, to be followed by further stabilization and linewidth reduction by locking to an ultra-high finesse ULE cavity. Finally, we point out that the use of a PPLN waveguide [3] instead of an enhancement cavity could further simplify the generation of yellow light.

**Acknowledgments**

This work was performed with the project "Space Optical Clocks" funded by the European Space Agency (ESA) and the Deutsches Zentrum für Luft- und Raumfahrt (DLR) (project 50 QT 0701). We thank A. Görlitz for helpful discussions and loan of equipment.